\documentclass[aps,twocolumn,showpacs]{revtex4}
\usepackage{amsmath,amsfonts,amssymb,wrapfig}
\def\beq{\begin{eqnarray}}
\def\eeq{\end{eqnarray}}
\def\mb#1{\mathbb{#1}}
\begin{document}
\title{Local symmetries of non-expanding horizons}
\author{Rudranil Basu}\email{rudranil@bose.res.in}
\author{Ayan Chatterjee}\email{achatterjee@imsc.res.in}
\author{Amit Ghosh}\email{amit.ghosh@saha.ac.in}
\affiliation{S. N. Bose National Centre for Basic Sciences, Block JD, Sector 3, Salt Lake, Kolkata 700098.\\
Institute of Mathematical Sciences, C.I.T. Campus, Taramani, Chennai 600113.\\
Saha Institute of Nuclear Physics, 1/AF Bidhannagar, Calcutta 700064.}
\begin{abstract}
Local symmetries of a non-expanding horizon has been investigated in the 1st
order formulation of gravity. When applied to a spherically symmetric isolated
horizon only a U(1) subgroup of the Lorentz group survives as residual local
symmetry that one can make use of in constructing an effective theory on the
horizon.
\end{abstract}
\pacs{04.70.Bw, 04.70.Dy }
\maketitle

In this note we explore local symmetries of a non-expanding horizon (NEH) in
the first order formulation of gravity. For a detailed definition of NEH see
\cite{ash1,aklrr}. For our present purpose it is sufficient to characterize NEH to
be a lightlike hypersurface $\Delta$ imbedded in spacetime such that the unique
(up to scaling by a function) lightlike, real vector field $l$ tangential to $\Delta$ is
expansion, shear and twist-free. Since $l$ is also normal to $\Delta$, it
is geodesic as well. These properties of $\Delta$ are independent of the scaling
of $l$ \cite{ash1,cg1}. Let us further assume that $\Delta$ is topologically equivalent to $\mb
S\times\mb R$ where $\mb S$ is a 2-sphere.

In the first order formulation Einstein's theory of gravity is invariant, apart
from diffeomorphisms, under the local Lorentz group. Here our specific interest
is primarily to find out the residual local symmetry of a NEH.
Then based on the residual gauge group we wish to propose an effective theory
on the horizon whose subsequent quantization would yield the quantum states of a
black hole. There is a recent upsurge of interest in such effective theories,
where an $SU(2)$ Chern-Simons theory has been proposed \cite{jpk} as the
effective quantum theory on the horizon in contrast to a $U(1)$ theory proposed
earlier \cite{ack,abck,abk}. In the canonical formulation of loop quantum gravity one
gauge fixes the full Lorentz group to its rotation subgroup $SU(2)$ and the
canonical theory reduces to a $SU(2)$ gauge theory. This is the main reason of
suspecting that a $SU(2)$ gauge theory (expectedly a topological one) may play a
role as the effective theory on the horizon in this case
\cite{smolin,Krasnov,Rovelli,km,abdv,bkm}. However, for a NEH our result goes in
favour of the $U(1)$ theory, as we exhibit below.

First, let us see how a NEH $\Delta$ reduces the local Lorentz symmetry. Being
expansion, shear and twist-free, certain Newman-Penrose coefficients
$\kappa_{\rm NP},\rho,\sigma$ vanish on $\Delta$; $\kappa_{\rm NP}$ vanishes
because the null-normal $l$ is a geodesic vector field, $\rho$ vanishes because
the expansion of $l$ vanishes and $\sigma$ vanishes because $l$ is shear-free
also. These conditions are satisfied only on $\Delta$. However, the
Newman-Penrose coefficients are sensitive to the local Lorentz transformations
\cite{Chandra}
\begin{align} &l\mapsto\xi l,n\mapsto\xi^{-1}n,m\mapsto m,\label{lor1}\\
&l\mapsto l,n\mapsto n,m\mapsto e^{i\theta}m,\label{lor2}\\
&l\mapsto l,n\mapsto n-cm-\bar c\bar m+c\bar cl,m\mapsto m-\bar
cl,\label{lor3}\\
&l\mapsto l-bm-\bar b\bar m+b\bar bl,n\mapsto n,m\mapsto m-\bar
bn,\label{lor4}
\end{align}
where $\xi,\theta,c,b$ are smooth functions on $\Delta$. Under (\ref{lor1}),
(\ref{lor2}) and (\ref{lor3}), $\kappa_{\rm NP},\rho,\sigma$
transform respectively as
\begin{align} &\kappa_{\rm NP}\mapsto\xi^2\kappa_{\rm NP},\;\rho\mapsto\xi\rho,\;\sigma\mapsto\xi\sigma\\ 
&\kappa_{\rm NP}\mapsto e^{i\theta}\kappa_{\rm NP},\;\rho\mapsto\rho,\;\sigma\mapsto e^{2i\theta} \sigma\\ 
&\kappa_{\rm NP}\mapsto\kappa_{\rm NP},\;\rho\mapsto\rho-c\,\kappa_{\rm NP},\;\sigma\mapsto\sigma-\bar c\,\kappa_{\rm NP}.\end{align}
Since they transform homogeneously, their vanishing remain invariant under
(\ref{lor1})-(\ref{lor3}). However, under (\ref{lor4}) they transform
inhomogeneously
\begin{align} &\kappa_{\rm NP}\mapsto\kappa _{\rm NP}-\bar b\rho-b\sigma+|b|^2\tau+2\bar b^2\alpha+2|b|^2\beta\nonumber\\
&\qquad~-2\bar b|b|^2\gamma-2\bar{b}\epsilon-\bar{b}|b|^2(\mu-\bar{\mu})+\bar{b}^2|b|^2\nu \nonumber\\ &\qquad~+\bar{b}^2\pi-\bar{b}^3 \lambda +D \bar{b} - b \delta \bar{b} - \bar{b} \bar{\delta} \bar{b} +|b|^2 \Delta \bar{b}\nonumber\\
&\rho\mapsto\rho - b \tau - 2 \bar{b} \alpha + 2 |b|^2 \gamma - \bar{b}|b|^2 \nu + \bar{b}^2 \lambda +\bar{ \delta} \bar{b}- b \Delta \bar{b}\nonumber\\
&\sigma\mapsto\sigma - \bar{b} \tau - 2\bar{b} \beta +2 \bar{b}^2 \gamma - \bar{b}^3 \nu + \bar{b}^2 \mu +\delta \bar{b} - \bar{b} \Delta \bar{b}\end{align}
where $D=\nabla_l$, $\Delta=\nabla_n$, $\delta=\nabla_m$ and $\bar\delta=\nabla_{\bar m}$. Clearly, the NEH boundary conditions 
are satisfied if and only if $b=0$.

The Lorentz matrices associated with the transformations
(\ref{lor1})-(\ref{lor3}) are respectively
\begin{align} \Lambda_{IJ}=&-\xi l_In_J-\xi^{-1}n_Il_J+2m_{(I}\bar
m_{J)},\label{L1}\\
\Lambda_{IJ}=&-2l_{(I}n_{J)}+(e^{i\theta}m_I\bar m_J+c.c.),\label{L2}\\
\Lambda_{IJ}=&-l_In_J-(n_I-cm_I-\bar cm_I+|c|^2l_I)l_J\nonumber\\
&+(m_I-\bar cl_I)\bar m_J+(\bar m_I-cl_I)m_J\label{L3}\end{align}
and the corresponding generators are respectively
\begin{align} &B_{IJ}=(\partial\Lambda_{IJ}/\partial\xi)_{\xi=1}=-2l_{[I}n_{J]},
\label{lbb}\\
&R_{IJ}=(\partial\Lambda_{IJ}/\partial\theta)_{\theta=0}=2im_{[I}\bar
m_{J]},\label{lbr}\\
&P_{IJ}=(\partial\Lambda_{IJ}/\partial{\rm Re}\,c)_{c=0}=2m_{[I}l_{J]}+2\bar
m_{[I}l_{J]},\label{lbp}\\
&Q_{IJ}=(\partial\Lambda_{IJ}/\partial{\rm Im}\,c)_{c=0}=2im_{[I}l_{J]}-2i\bar
m_{[I}l_{J]},\label{lbq}\end{align}
where $B,R$ generate (\ref{lor1}) and (\ref{lor2}) respectively and $P,Q$
generate (\ref{lor3}). A straightforward calculation gives their Lie brackets
\begin{align} &[R,B]=0,\quad [R,P]=Q,\quad [R,Q]=-P,\nonumber\\
&[B,P]=P,\quad [B,Q]=Q,\quad [P,Q]=0,\end{align}
where $[R,B]_{IJ}=R_{IK}B^K{}_J-B_{IK}R^K{}_J$ and so on. This is the Lie
algebra of $ISO(2)\ltimes\mb R$ where the symbol $\ltimes$ stands for the semidirect
product; $R,P,Q$ generate $ISO(2)$ and $B$ generates $\mb R$.

Clearly, the NEH boundary conditions are invariant only under a subgroup of
the local Lorentz group. We should keep note of the fact that the group
$ISO(2)\ltimes\mb R$ is non-semisimple; its Cartan-Killing metric $K$ is
doubly degenerate
\beq K=\bordermatrix{&R&B&P&Q\cr &-2&0&0&0\cr &0&2&0&0\cr &0&0&0&0\cr
&0&0&0&0}.\label{ckm}\eeq

Let us consider the Palatini connection $\mb A_{IJ}$ and in the interior of
the spacetime let us expand $\mb A_{IJ}$ in the internal Lorentz basis
\begin{align} \mb A_{IJ}=&-2\mb Wl_{[I}n_{J]}+2\mb Vm_{[I}\bar m_{J]}+2(\bar{\mb N}n_{[I}m_{J]}+c.c.)\nonumber\\
 &+2(\bar{\mb U}l_{[I}m_{J]}+c.c.)\label{balc}\end{align}
where $\mb W,\mb V,\mb N,\mb U$ are connection 1-forms; as defined, $\mb W$ is
real, $\mb V$ is imaginary and $\mb N,\mb U$ are complex (in all, there are six
of them associated with the six generators). For the rest of our analysis we
will fix an internal Lorentz frame for which $l_I,n_I,m_I,\bar m_I$ are
constants. However, our results will be unaffected by such a choice.

The pull-back of the Palatini connection to the NEH $\Delta$ is of the form
\beq A_{IJ}\triangleq-2Wl_{[I}n_{J]}+2Vm_{[I}\bar m_{J]}+2(\bar Ul_{[I}m_{J]}+c.c.)
\label{palc}\eeq
where $W,V,U$ are respectively the pull-backs of $\mb W,\mb V,\mb U$. Clearly,
the 1-form $N$, which is the pull-back of $\mb N$, vanishes on $\Delta$ by
the NEH boundary conditions. Proof: The simplest way to show this is to relate
the connection 1-forms to the Newman-Penrose coefficients (the constant
$l_I,n_I,m_I,\bar m_I$ basis simplifies these relations):
\begin{align} &\mb
W=\!-(\gamma+\bar\gamma)l-(\epsilon+\bar\epsilon)n+(\alpha+\bar\beta)m+(\bar\alpha+
\beta)\bar m\label{wnp}\\
&\mb
V=-(\gamma-\bar\gamma)l-(\epsilon-\bar\epsilon)n+(\alpha-\bar\beta)m+(\beta-\bar
\alpha)\bar m\label{vnp}\\
&\mb U=-\bar\nu l-\bar\pi n+\bar\mu m+\bar\lambda\bar m\label{unp}\\
&\mb N=\tau l+\kappa_{\rm NP}n-\rho m-\sigma\bar m.\label{nnp}\end{align}
So only four independent connection 1-forms $W,V,U$ survive on $\Delta$. This is
consistent with our earlier result that the residual gauge group on $\Delta$ is
$ISO(2)\ltimes\mb R$ that has only four generators. However, below we present an
independent analysis for the connection to prove this.

Under the local Lorentz transformations (\ref{lor1})-(\ref{lor4}) the Palatini
connection (\ref{balc}) transform as
\beq \mb A_{IJ}\mapsto\Lambda_I{}^K\mb A_{KL}\Lambda_J{}^L+\Lambda_{IK}d\Lambda_J{}^K\label{ptrans}\eeq
where $\Lambda_{IJ}$ are the associated Lorentz matrices (\ref{L1})-(\ref{L3})
for (\ref{lor1})-(\ref{lor3}) and for (\ref{lor4})
\begin{align} \Lambda_{IJ}=&-(l_I-bm_I-\bar b\bar m_I+b\bar bn_I)n_J-n_Il_J\nonumber\\
&+(m_I-\bar bn_I)\bar m_J+(\bar m_I-bn_I)m_J.\label{L4}\end{align}
A lengthy but straightforward calculation shows that under the Lorentz
transformations (\ref{L1})-(\ref{L3}) the connection 1-forms transform as
\begin{align} &\mb W\mapsto\mb W-d\ln\xi,\mb V\mapsto\mb V,\mb U\mapsto\xi\mb
U,\mb N\mapsto \xi^{-1}\mb N.\label{C1}\\
&\mb W\mapsto\mb W,\mb V\mapsto\mb V-id\theta,\mb U\mapsto e^{-i\theta}\mb U,\mb
N\mapsto e^{-i\theta}\mb N.\label{C2}\\
&\mb W\mapsto\mb W-c\mb N-\bar c\bar{\mb N},\mb V\mapsto\mb V-c\mb N+\bar
c\bar{\mb N},\nonumber\\ &\qquad~~\mb U\mapsto\mb U-d\bar c+\bar c(\mb W-\mb
V)-\bar c^2\bar{\mb N},\mb N\mapsto\mb N.\label{C3}\end{align}
Since $\mb N$ transforms homogeneously, its pull-back $N\triangleq 0$ in one
frame implies that it vanishes in all Lorentz frames related by
(\ref{L1})-(\ref{L3}). However, under (\ref{L4}), the connection 1-forms
transform as
\begin{align} &\mb W\mapsto\mb W+b\mb U+\bar b\bar{\mb U},\mb V\mapsto\mb V-b\mb U+\bar b\bar{\mb U}\nonumber\\
 &\mb U\mapsto\mb U,\mb N\mapsto\mb N+d\bar b-\bar b(\mb W+\mb V)-\bar b^2\bar{\mb U}.\label{C4} \end{align}
Clearly, in this case $N\triangleq 0$ if and only if $b$ satisfies the equation
$db\triangleq b(W-V+b\bar U)=:bY$ where $Y$ is a 1-form. This equation has
a nontrivial solution if and only if $Y$ is a closed 1-form. However, we show
that the equation admits only the trivial solution, $b=0$. Proof: Since $b$ is a
constant in the phase space of a NEH, it is sufficient to show that $Y$ is not
closed for one specific NEH. Consider for example the event horizon of the Schwarzschild
solution. In units $G=1$ and in advanced Eddington-Finkelstein coordinates
\beq W=\frac{1}{4M}\,dv,\;U=-\frac{1}{\sqrt 2}(d\theta+i\sin\theta\,d\phi),\nonumber\\
V=-i\cos\theta\,d\phi.\eeq
As a result, $dV$ and $dU$ are proportional to the 2-sphere area 2-form
and $dW=0$. However, since $Y$ depends on $b$, one can ask is there any $b$ for which
$dY\triangleq 0$? The answer is explicitly verifiable and one easily finds
that $dY\triangleq 0$ if and only if $b=0$. Since $Y$ is not closed, acting
$d$ once more on the equation $db=bY$ one gets
\beq 0=db\wedge Y+b\,dY=bY\wedge Y+b\,dY=b\,dY,\eeq
which yields the unique solution $b=0$. This shows that the connection
(\ref{palc}) is indeed an $ISO(2) \ltimes\mb R$ connection. Here we wish to
remark that one could also arrive at (\ref{C1})-(\ref{C4}) directly using the
relations (\ref{wnp})-(\ref{nnp}) and the appropriate Lorentz transformations
of the Newman-Penrose coefficients \cite{Chandra}.

It is to be noted that unlike the Palatini connection, the H\"olst connection
$\mb H_{IJ}:=\mb A_{IJ}-\frac{1}{2}\gamma_B\epsilon_{IJ}{}^{KL}\mb A_{Kl}$,
where $\gamma_B$ is the Barbero-Immirzi parameter, does not transform as a
connection under any of the local Lorentz transformations
(\ref{lor1})-(\ref{lor4}).

For later convenience we expand (\ref{palc}) in the basis
(\ref{lbb})-(\ref{lbq}) of the Lie algebra $\mathfrak{iso}(2)\ltimes\mb R$:
\beq A_{IJ}=2A_BB_{IJ}+2A_RR_{IJ}+2A_PP_{IJ}+2A_QQ_{IJ}\eeq
where $2A_B=W$, $2A_R=-iV$, $2A_P=-{\rm Re}\,U$ and $2A_Q={\rm Im}\,U$. The
connection 1-forms $A_B,A_R,A_P,A_Q$ will turn out to be more useful in the
context of an effective theory on the horizon.

Let us now turn our attention to the symplectic structures. The H\"olst action
\cite{holst} gives rise to the symplectic current 3-form (in units of $4\pi
G\gamma_B=1$ and $\mb E^I$ is the spacetime tetrad 1-form)
\beq \mb J(\delta_1,\delta_2)=-\frac{1}{4}\delta_1(\mb E^I\wedge\mb
E^J)\wedge\delta_2\mb H_{IJ}-(1\leftrightarrow 2).\eeq
Its pull-back to $\Delta$ gives the boundary symplectic current. For
simplicity, we take $\Delta$ for the rest of our analysis to be a spherically
symmetric isolated horizon. A straightforward calculation gives the
pull-back current $J$
\beq J(\delta_1,\delta_2)\triangleq\frac{1}{2}\delta_{1}{}^{2}{\epsilon}
\wedge\delta_2(iV+\gamma_BW)-(1\leftrightarrow 2)\label{sympc}\eeq
where ${}^{2}{\epsilon}$ is the area 2-form of some spherical cross-section of
$\Delta$. In the derivation of the symplectic current it is sufficient
to assume that the spherical cross-section foliates $\Delta$ and is not
necessarily a geometric 2-sphere. However, for the rest of our analysis
we will restrict ourselves to the unique foliation of $\Delta$ in which
each leaf is a geometric 2-sphere; this is possible if and only if the
isolated horizon $\Delta$ is spherically symmetric. For such a horizon
with a fixed area $\mathcal{A}=\int{}^{2}{\epsilon}$ the 1-form $W$
is closed and $dV$ is proportional to ${}^{2}{\epsilon}$ \cite{abf,ayan}
\beq dW\triangleq 0,\quad dV\triangleq\frac{4\pi
i}{\mathcal{A}}\,{}^{2}{\epsilon}\label{1forms}\eeq
where $d$ is the exterior derivative intrinsic to $\Delta$.
Using (\ref{1forms}) we find that the symplectic current 3-form is exact on
$\Delta$
\begin{align} &J(\delta_1,\delta_2)\triangleq
dj(\delta_1,\delta_2)\;\mbox{where}\notag\\
&j(\delta_1,\delta_2)=-\frac{\mathcal{A}}{8\pi}\delta_1(iV+\gamma_BW)\wedge\delta_2
(iV+\gamma_BW).\label{sympcd}\end{align}
It is to be noted that in the $\mathfrak{iso}(2)\ltimes\mb R$ basis
the 1-form $iV+\gamma_BW=-2(A_R-\gamma_BA_B)$. This gives a boundary symplectic
structure (putting back $4\pi G\gamma_B=1$)
\beq \Omega(\delta_1,\delta_2)=-\frac{\mathcal{A}}{8\pi^2G\gamma_B}\int_{\mb
S}\delta_1A_{\rm CS}\wedge\delta_2A_{\rm CS}\label{symp}\eeq
where $\mb S$ is the unique spherical cross-section of $\Delta$ and $A_{\rm
CS}=A_R-\gamma_BA_B$.

The form (\ref{symp}) suggests that on a spherically symmetric isolated horizon
one can take the effective boundary theory as a $U(1)$ Chern-Simons theory. Two
distinct cases of $U(1)$ arise: {\tt i)} If either the pull-back of $A_B$
vanishes on $\mb S$ \cite{Lewandowski} or one restricts the gauge freedom (\ref{lor1}) to a
constant class ($\xi=$ constant, as has been the original choice \cite{ack})
then one gets a compact $U(1)$, {\tt ii)} In general, if no restrictions are
imposed, then one gets a noncompact $U(1)$.

\begin{acknowledgments}
We thank Abhay Ashtekar for correspondence, Parthasarathi Majumdar and Ramesh Kaul at
the initial stage of collaboration and Parthasarathi Mitra for discussion. RB thanks
Council for Scientific and Industrial Research (CSIR), India, for support through the
SPM Fellowship SPM-07/575(0061)/2009-EMR-I.
\end{acknowledgments}


\begin{thebibliography}{99}
\bibitem{ash1}
A.~Ashtekar, S.~Fairhurst and B.~Krishnan,
  Phys.\ Rev.\  D {\bf 62}, 104025 (2000)
  [arXiv:gr-qc/0005083].
\bibitem{aklrr}
A.~Ashtekar and B.~Krishnan,
  Living Rev.\ Rel.\  {\bf 7}, 10 (2004)
  [arXiv:gr-qc/0407042]
\bibitem{cg1}
  A.~Chatterjee and A.~Ghosh,
  Class.\ Quant.\ Grav.\  {\bf 23}, 7521 (2006)
  [arXiv:gr-qc/0603023].
\bibitem{jpk}
  J.~Engle, A.~Perez and K.~Noui,
  arXiv:0905.3168 [gr-qc].
\bibitem{abck}
  A.~Ashtekar, J.~Baez, A.~Corichi and K.~Krasnov,
  Phys.\ Rev.\ Lett.\  {\bf 80}, 904 (1998)
  [arXiv:gr-qc/9710007].
\bibitem{ack}A.~Ashtekar, A.~Corichi and K.~Krasnov,
  Adv.\ Theor.\ Math.\ Phys.\  {\bf 3}, 419 (2000)
  [arXiv:gr-qc/9905089].
\bibitem{abk}
  A.~Ashtekar, J.~C.~Baez and K.~Krasnov,
  Adv.\ Theor.\ Math.\ Phys.\  {\bf 4}, 1 (2000)
  [arXiv:gr-qc/0005126].
\bibitem{smolin}
  L.~Smolin,
  J.\ Math.\ Phys.\  {\bf 36}, 6417 (1995)
  [arXiv:gr-qc/9505028].
\bibitem{Krasnov}
  K.~V.~Krasnov,
  Phys.\ Rev.\  D {\bf 55}, 3505 (1997)
  [arXiv:gr-qc/9603025].

\bibitem{Rovelli}C.~Rovelli,
  Phys.\ Rev.\ Lett.\  {\bf 77}, 3288 (1996)
  [arXiv:gr-qc/9603063].

\bibitem{km} R.~K.~Kaul and P.~Majumdar,
  Phys.\ Rev.\ Lett.\  {\bf 84}, 5255 (2000)
  [arXiv:gr-qc/0002040].
\bibitem{abdv}
  I.~Agullo, G.~J.~Fernando Barbero, E.~F.~Borja, J.~Diaz-Polo and E.~J.~S.~Villasenor,
  Phys.\ Rev.\  D {\bf 80}, 084006 (2009)
  [arXiv:0906.4529 [gr-qc]].

\bibitem{bkm}R.~Basu, R.~K.~Kaul and P.~Majumdar,
  arXiv:0907.0846 [gr-qc].
\bibitem{Chandra}J.~M.~Stewart,
{\it  Cambridge University Press, Cambridge, 1990}; S.~Chandrasekhar,
{\it  Oxford, UK: Clarendon (1992) 646 p.}
\bibitem{holst} S.~H\"olst,
        Phys.\ Rev.\  D {\bf 53}, 5966 (1996)
        [arXiv:gr-qc/9511026].
\bibitem{abf}A.~Ashtekar, C.~Beetle and S.~Fairhurst,
  Class.\ Quant.\ Grav.\  {\bf 17}, 253 (2000)
  [arXiv:gr-qc/9907068].
\bibitem{ayan} A.~Chatterjee and A.~Ghosh,
  Phys.\ Rev.\  D {\bf 80}, 064036 (2009)
  [arXiv:0812.2121 [gr-qc]].
\bibitem{Lewandowski}J.~Lewandowski,
  Class.\ Quant.\ Grav.\  {\bf 17}, L53 (2000)
  [arXiv:gr-qc/9907058].
\end{thebibliography}
\end{document}